%% file: main.tex
\documentclass[acmtog,screen,balance=true]{acmart}

\usepackage{booktabs} 
\usepackage{overpic_a}
\usepackage{subcaption}
\usepackage{bm}
\usepackage{array}
\usepackage{textcomp}
\usepackage{caption}
\usepackage{xspace}
\usepackage{multirow}
\usepackage{url}
\usepackage{wrapfig}
\usepackage{graphicx}
\usepackage{xcolor}
\usepackage{siunitx}
\usepackage{hyperref}

\author{Songyin Wu}
\thanks{The project was completed during Songyin Wu’s internship at Meta Reality Labs Research.}
\email{s_wu975@ucsb.edu}
\orcid{0009-0009-9581-2506}
\affiliation{%
  \institution{University of California Santa Barbara}
  \country{USA}
}

\author{Zhaoyang Lv}
\email{zhaoyang@meta.com}
\orcid{0000-0002-7788-9982}
\affiliation{%
  \institution{Meta Reality Labs Research}
  \country{USA}
}

\author{Yufeng Zhu}
\email{yufengzhu@meta.com}
\orcid{0009-0003-8716-201X}
\affiliation{%
  \institution{Meta Reality Labs Research}
  \country{USA}
}

\author{Duncan Frost}
\email{frost@meta.com}
\orcid{0000-0002-5948-3452}
\affiliation{%
  \institution{Meta Reality Labs Research}
  \country{United Kingdom}
}

\author{Zhengqin Li}
\email{zhl@meta.com}
\orcid{0000-0003-0868-2141}
\affiliation{%
  \institution{Meta Reality Labs Research}
  \country{USA}
}

\author{Ling-Qi Yan}
\email{lingqi@cs.ucsb.edu}
\orcid{0000-0002-9379-094X}
\affiliation{%
  \institution{University of California Santa Barbara}
  \country{USA}
}

\author{Carl Ren}
\email{carlren@meta.com}
\orcid{0009-0000-0942-2929}
\affiliation{%
  \institution{Meta Reality Labs Research}
  \country{USA}
}

\author{Richard Newcombe}
\email{newcombe@meta.com}
\orcid{0009-0004-9091-8989}
\affiliation{%
  \institution{Meta Reality Labs Research}
  \country{USA}
}

\author{Zhao Dong}
\email{zhaodong@meta.com}
\orcid{0000-0002-9026-6886}
\affiliation{%
  \institution{Meta Reality Labs Research}
  \country{USA}
}

\usepackage[ruled]{algorithm2e} 

\SetAlFnt{\small}
\SetAlCapFnt{\small}
\SetAlCapNameFnt{\small}
\SetAlCapHSkip{0pt}

\copyrightyear{2025}
\acmYear{2025}
\acmConference[SIGGRAPH Conference Papers '25]{Special Interest Group on Computer Graphics and Interactive Techniques Conference Conference Papers }{August 10--14, 2025}{Vancouver, BC, Canada}
\acmBooktitle{Special Interest Group on Computer Graphics and Interactive Techniques Conference Conference Papers (SIGGRAPH Conference Papers '25), August 10--14, 2025, Vancouver, BC, Canada}
\acmDOI{10.1145/3721238.3730659}
\acmISBN{979-8-4007-1540-2/2025/08}

\begin{document}



\title{Monocular Online Reconstruction with Enhanced Detail Preservation}

\begin{abstract}
\input{abstract}
\end{abstract}

\input{Figures/teaser}

\maketitle
\renewcommand{\shortauthors}{Wu et al.}

\input{introduction}
\input{related}
\input{methods}
\input{experiments}
\input{conclusion}

\begin{acks}
We would like to thank Chieh Hubert Lin, Lei Xiao, Numair Khan, Chris Sweeney, Julian Straub, and Carl Marshall for their valuable discussions on the project. We also thank the anonymous reviewers for their insightful feedback. The pseudocode template in the supplementary is from \citet{zhou2024unifiedgaussianprimitivesscene}. The project was completed during Songyin Wu’s internship at Meta Reality Labs Research. Ling-Qi Yan is supported by gift funds from Meta, Adobe, Linctex, and XVerse.
\end{acks}

\bibliographystyle{ACM-Reference-Format}
\bibliography{paper}

\newpage

\input{Figures/Aria_comparison}
\input{Figures/image_only}

\end{document}

%% file: abstract.tex
We propose an online 3D Gaussian-based dense mapping framework for photorealistic details reconstruction from a monocular image stream. Our approach addresses two key challenges in monocular online reconstruction: distributing Gaussians without relying on depth maps and ensuring both local and global consistency in the reconstructed maps. To achieve this, we introduce two key modules: the \textit{Hierarchical Gaussian Management Module} for effective Gaussian distribution and the \textit{Global Consistency Optimization Module} for maintaining alignment and coherence at all scales. In addition, we present the \textit{Multi-level Occupancy Hash Voxels} (MOHV), a structure that regularizes Gaussians for capturing details across multiple levels of granularity. MOHV ensures accurate reconstruction of both fine and coarse geometries and textures, preserving intricate details while maintaining overall structural integrity. Compared to state-of-the-art RGB-only and even RGB-D methods, our framework achieves superior reconstruction quality with high computational efficiency. Moreover, it integrates seamlessly with various tracking systems, ensuring generality and scalability. Project page: \url{https://poiw.github.io/MODP/}.

%% file: Figures/teaser.tex
\begin{teaserfigure}
    \centering
    \begin{overpic}[width=\linewidth]{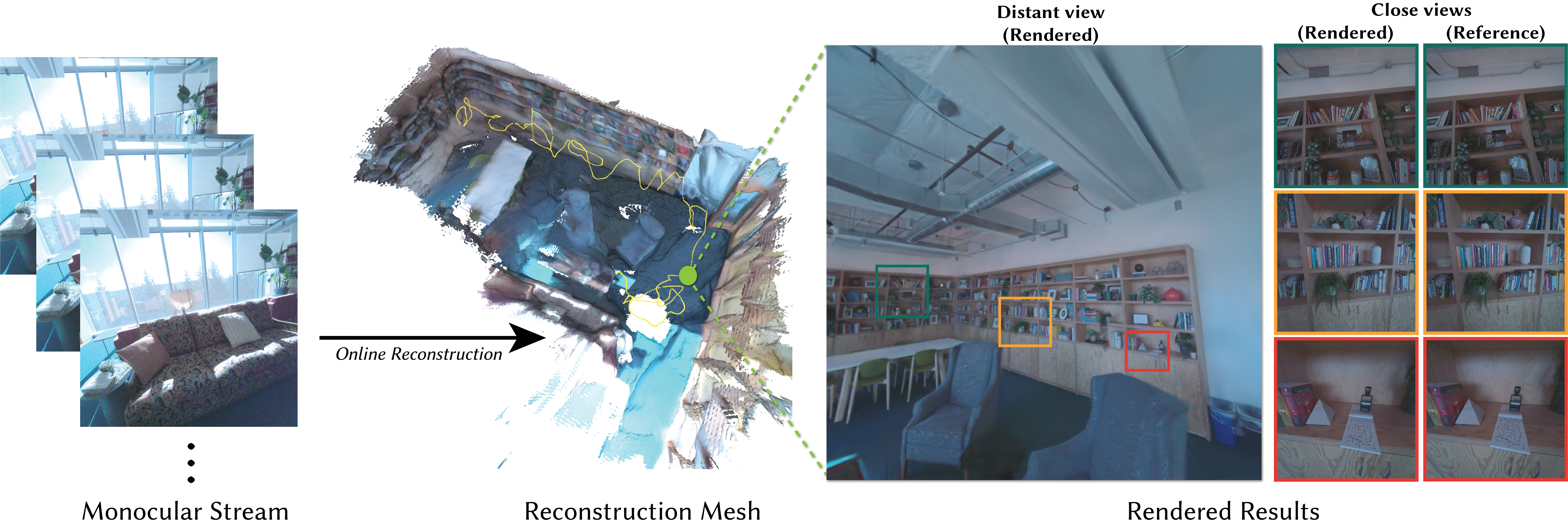}
    \end{overpic}
    \caption{Our pipeline processes a stream of monocular RGB images to reconstruct scenes with immediate feedback. Our method produces high-quality photorealistic maps with detailed reconstruction across multiple levels. The middle image illustrates our reconstructed mesh, and the right image showcases the rendered results of our reconstructed map, which captures high-quality details at both coarse and fine levels.}
    \label{fig:teaser}
\end{teaserfigure}

%% file: introduction.tex
\section{Introduction}

Online dense reconstruction, which generates environment models from a continuous stream of input images, is a fundamental challenge in robotics, computer vision, and computer graphics. It forms the cornerstone of interactive environmental understanding and interaction, enabling a wide range of applications such as augmented and virtual reality (AR/VR), robotics, and the emerging field of spatial AI. By processing sensor data streams to interactively reconstruct environments, it provides immediate feedback to the scanner, facilitating downstream tasks like scene understanding, active reconstruction, and more.

Previous works have explored various map representations, including point clouds~\cite{du2011interactive}, surfels~\cite{keller2013real, whelan2015elasticfusion}, and signed distance functions~\cite{newcombe2011kinectfusion}, to achieve high-quality geometry reconstruction in Simultaneous Localization and Mapping (SLAM) systems. However, these methods often fall short in reconstructing photorealistic appearances due to the limitations of their map representations, as their primary focus is on geometric accuracy rather than visual realism. The recent success of neural radiance fields (NeRFs) in view synthesis~\cite{mildenhall2020nerf} has paved the way for photorealistic scene reconstruction. NeRF-based methods~\cite{sucar2021imap, zhu2022nice, yang2022vox, johari2023eslam, wang2023co, sandstrom2023point, zhu2024nicer}, when integrated into SLAM systems, mark a significant breakthrough in enhancing reconstructed maps with highly realistic appearances. However, these methods face challenges in achieving interactive frame rates for both reconstruction and rendering due to the computational demands of ray marching in volumetric rendering. Additionally, their high memory requirements make it difficult to scale effectively to large scenes. In contrast, 3D Gaussian representations~\cite{kerbl20233d} model scenes as discrete Gaussian distributions, offering dramatically faster rendering and optimization speeds. This allows reconstructed scenes to be visualized at real-time frame rates. Recent works~\cite{Matsuki:Murai:etal:CVPR2024, yan2024gs, yugay2023gaussian, keetha2024splatam,ha2025rgbd,huang2024photo,sandstrom2024splat,zhang2024hi, bai2024rpslam} have demonstrated that 3D Gaussian-based SLAM can produce high-quality reconstruction maps, achieving superior rendering quality compared to earlier non-radiance-field-based methods.

Dense SLAM systems can process various types of input streams. 3D Gaussian-based dense SLAMs with RGB-D inputs~\cite{Matsuki:Murai:etal:CVPR2024, peng2024rtg, ha2025rgbd, wang2024ogmap} achieve high-quality scene reconstruction, as the availability of accurate depth data allows Gaussians to be initialized near optimal locations. This facilitates rapid convergence with precise geometry and appearance. However, the reconstruction quality deteriorates significantly when relying solely on color frames (monocular input). Poor initialization of Gaussians often leads to independent optimization getting trapped in local minima, resulting in artifacts such as floaters and blurriness.
In contrast to offline methods~\cite{kerbl20233d}, which benefit from accurate camera poses and global optimization of the entire scene, incremental reconstruction in online SLAM systems faces challenges such as limited computational resources and the absence of global information. These issues lead to inconsistencies and lower-quality global maps. As a result, achieving photorealistic online reconstruction without depth maps remains a significant challenge.

In this paper, we aim to achieve photorealistic reconstruction using only monocular RGB frames, addressing two key challenges: the lack of dense depth maps and the need to produce globally consistent maps at interactive frame rates. Our key insight lies in controlling the distribution of Gaussians in world space based on error maps and feature complexity in image space. Furthermore, we design a multi-level occupancy hash voxel structure to regulate the distribution across different levels, ensuring coarse and fine details are recovered. In addition, we propose a view selection strategy that balances the reconstruction of newly observed local regions with the preservation of historical global maps, effectively avoiding local minima during optimization.

We evaluate our pipeline on standard datasets, including TUM~\cite{sturm2012benchmark} and Replica~\cite{straub2019replica}, where it demonstrates superior reconstruction quality compared to previous monocular baselines and surpasses most RGB-D baselines. To further validate its capability to handle scenes with varying levels of detail for scenes in different scales, we capture additional indoor and outdoor sequences with complex geometries using Aria glass~\cite{engel2023project}. Moreover, our proposed mapping system is designed to be compatible with various tracking systems, highlighting its versatility. We showcase the generality of our approach by integrating it with different tracking systems, such as those in~\cite{campos2021orb, engel2023project}, demonstrating its scalability and effectiveness.

%% file: related.tex
\section{Related works}
\input{Figures/overview}
\subsection{Classic Dense Visual SLAM}
Over the last decade, dense visual SLAM-based 3D scene reconstruction has been a prominent research focus. For a comprehensive overview, readers are referred to detailed state-of-the-art surveys~\cite{macario2022comprehensive, zollhofer2018state, fuentes2015visual} and foundational theses~\cite{richard2012SLAMthesis}. Significant advancements in online 3D scene reconstruction have been achieved in RGB-D dense SLAM, employing diverse map representations such as point clouds~\cite{du2011interactive}, Hermite radial basis functions~\cite{xu2022hrbf}, surfels~\cite{cao2018real,keller2013real,whelan2015elasticfusion}, and truncated signed distance functions (TSDFs)~\cite{chen2013scalable, dai2017bundlefusion, huang2021real, newcombe2011kinectfusion, niessner2013real, zhang2015online}. For instance, ElasticFusion~\cite{whelan2015elasticfusion} models scenes as collections of surfels, leveraging surfel-rendered depth and color for high-quality real-time tracking. TSDF-based BundleFusion~\cite{dai2017bundlefusion} reconstructs large-scale scenes in real time through dynamic surface reintegration, achieving globally consistent 3D maps. DI-Fusion~\cite{huang2021di} incorporates scene priors by encoding local geometry and modeling uncertainty using deep neural networks. While these methods focus primarily on geometric reconstruction, our approach simultaneously addresses surface reconstruction and photorealistic rendering, bridging the gap between accurate geometry and visually realistic output. Moreover, these prior methods rely heavily on depth maps to achieve high-quality reconstruction. In contrast, our approach operates using only a monocular RGB stream, making it more versatile and accessible.

\subsection{NeRF-based Dense Visual SLAM}
Building upon the remarkable success of neural radiance fields (NeRF)~\cite{mildenhall2020nerf}, recent works have integrated NeRF with RGB-D and RGB-only dense SLAM systems. In RGB-D SLAM, iMap~\cite{sucar2021imap} pioneers NeRF SLAM by using a single MLP to represent the scene. NICE-SLAM~\cite{zhu2022nice} introduces hierarchical feature grids decoded with pre-trained MLPs, while Vox-Fusion~\cite{yang2022vox} represents scenes as voxel-based neural implicit surfaces stored in octrees. State-of-the-art methods, such as ESLAM~\cite{johari2023eslam} and Co-SLAM~\cite{wang2023co}, adopt multi-resolution hierarchical structures to balance quality and performance, using feature grids and hash grids, respectively. Point-SLAM~\cite{sandstrom2023point} takes an alternative approach, leveraging neural point clouds with volumetric rendering and feature interpolation. For RGB-only SLAM, NICER-SLAM~\cite{zhu2024nicer} enhances accuracy and robustness by incorporating additional supervision signals, such as monocular geometric cues and optical flow, to jointly optimize camera poses and hierarchical neural implicit maps. While these methods deliver impressive results, their reliance on computationally intensive volumetric rendering limits interactive or real-time performance for online reconstruction of real-world scenes. Furthermore, their high memory requirements make reconstructing large-scale scenes impractical. In contrast, our method enables interactive reconstruction of large-scale scenes, offering significantly higher speeds and reduced memory costs.

\subsection{3D Gaussian-based Dense Visual SLAM}
Recently, 3D Gaussians~\cite{kerbl20233d} have gained traction as an efficient alternative for map representation in RGB-D and RGB-only dense SLAM systems. Traditional 3D Gaussian optimization, typically performed offline, requires several minutes to complete. To enable online reconstruction in RGB-D SLAM, methods such as~\cite{yan2024gs, yugay2023gaussian, wang2024ogmap} introduce novel Gaussian seeding and optimization strategies for sequential input streams. SplaTAM~\cite{keetha2024splatam} incorporates Gaussian-based representations with silhouette-guided optimization via differentiable rendering. MonoGS~\cite{Matsuki:Murai:etal:CVPR2024} extends Gaussian representations for accurate tracking, mapping, and high-quality rendering in both RGB-D and RGB-only scenarios. More recently, RTG-SLAM~\cite{peng2024rtg} introduces a compact Gaussian representation with a highly efficient on-the-fly optimization scheme for RGB-D inputs, achieving real-time online scene reconstruction. GS-ICP SLAM~\cite{ha2025rgbd} combines Generalized Iterative Closest Point (G-ICP) with 3D Gaussian Splatting (3DGS) to further enhance real-time RGB-D SLAM performance. For RGB-only SLAM, Photo-SLAM~\cite{huang2024photo} utilizes a Gaussian-Pyramid training approach to improve mapping with multi-level features. Splat-SLAM~\cite{sandstrom2024splat} dynamically adapts to keyframe pose and depth updates by deforming the 3D Gaussian map, ensuring globally optimized tracking and enhanced reconstruction accuracy. HI-SLAM2~\cite{zhang2024hi} combines monocular priors with learning-based dense SLAM to improve geometry estimation, achieving significant advancements in quality and performance for RGB-only SLAM.
Some other concurrent works~\cite{bai2024rpslam, tianci2025scaffoldslam, hu2024msoslamc, feng2024cartgs} also attempt to reconstruct high-quality scenes from a monocular stream; however, none of them show very fine detailed reconstruction. 
Compared to these existing methods, our approach aims to achieve an optimal balance between reconstruction quality and performance. Quality-wise, our method excels in preserving fine details, significantly outperforming all RGB-only approaches and surpassing most RGB-D-based methods on standard datasets~\cite{sturm2012benchmark, straub2019replica}. Performance-wise, it delivers interactive online reconstruction.

%% file: Figures/overview.tex
\begin{figure*}[tbh]
    \centering
    \begin{overpic}[width=\linewidth]{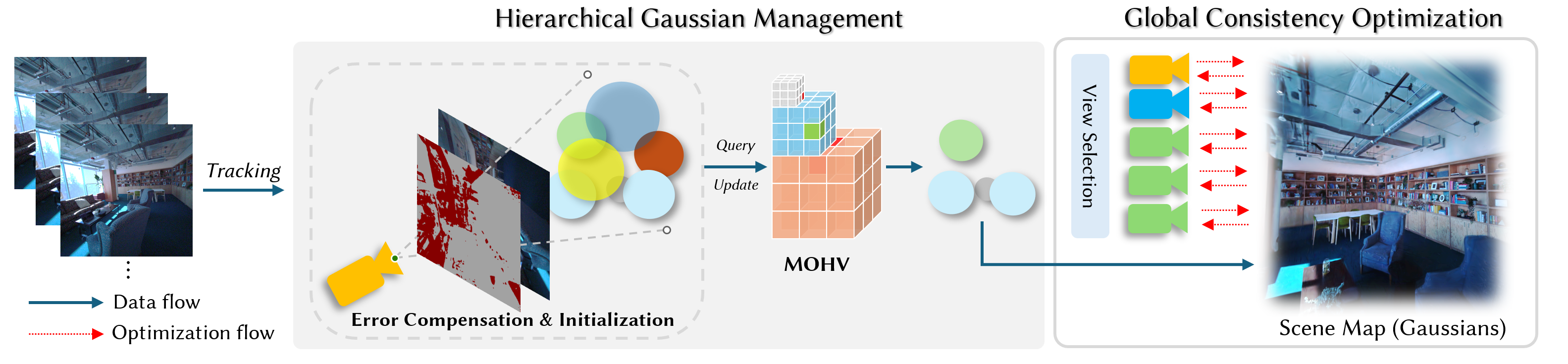}
    \end{overpic}
    
    \caption{The figure illustrates an overview of our method. Our approach takes inputs from the tracking system and generates initial Gaussians based on regions with high geometric or texture complexity and significant errors (Sec.~\ref{sec:densification}). Next, the MOHV module (Sec.~\ref{sec:MOHV}) removes redundant Gaussians while preserving a high-quality map. Finally, the global consistency optimization module optimizes the Gaussians to produce a globally consistent map with details across various levels.
    }
    \label{fig:overview}
\end{figure*}

%% file: methods.tex
\section{Methods}
The high-level ideas of our design are based on two key observations: the \emph{importance of good Gaussian initialization} and the \emph{need for global consistency optimization}. With only monocular input, achieving a well-initialized Gaussian distribution through regularization is crucial to decrease artifacts, including floater and over-blurriness, where we propose our \emph{Hierarchical Gaussian Management} module~(Sec.~\ref{sec:hgm}) with a densification strategy (Sec.~\ref{sec:densification}) and a pruning mechanism through MOHV (Sec.~\ref{sec:MOHV}). Building on this solid initialization, we propose \emph{Global Consistency Optimization} (Sec.~\ref{sec:global_opt}) to balance local rapid convergence with global consistency. An overview of our method is presented in Fig.~\ref{fig:overview}. 

\subsection{Tracking}
Our method emphasizes a high-quality mapping system while ensuring compatibility with various tracking systems.  
We use ORB-SLAM3~\cite{campos2021orb} as an example tracking system in our pipeline, where it provides online camera poses $c_i$ and sparse world space feature points $P_i$ to our system.
Additional results using different tracking systems will be presented in Sec.~\ref{sec:aria_tracking}, showcasing the generalization and robustness of our mapping system. Note that we do not discuss the quality of different tracking systems, as our focus is on the mapping system.   

\subsection{Hierarchical Gaussian Management} \label{sec:hgm}
Our pipeline uses 3D Gaussians~\cite{kerbl20233d} as scene representations, where the initialization and distribution of 3D Gaussians are critical, as each Gaussian is optimized independently. Previous RGB-D SLAM works~\cite{Matsuki:Murai:etal:CVPR2024, peng2024rtg, ha2025rgbd} rely on depth maps to precisely initialize Gaussians, achieving higher reconstruction quality compared to methods using only RGB inputs~\cite{Matsuki:Murai:etal:CVPR2024, huang2024photo, sandstrom2024splat, zhang2024hi}. Instead, we propose a hierarchical Gaussian management module to avoid issues such as floaters and missing details caused by local minima. This module computes Gaussian and camera scales, strategically distributing Gaussians to geometry-complex and high-error regions to capture geometry and texture details better. Furthermore, to address redundancy or insufficiency in Gaussian placement across varying levels of scene detail, we propose a Multi-level Occupancy Hash Voxel (MOHV) structure. MOHV dynamically regulates Gaussian density at different scales, enabling high-quality reconstruction of fine details while maintaining computational efficiency.

\subsubsection{Camera and Gaussian Scales} \label{sec:cam_scale}
Gaussian scales are critical in the optimization process: excessively large scales cause over-blurred results, while excessively small scales hinder convergence. To effectively capture details at varying levels with appropriate scales, we define the scales of both the camera and Gaussians in world space. Specifically, given the camera's focal length $f_i$ and the pixel's depth value $d_t$ (from the tracker), the Gaussian size corresponding to such a pixel is calculated using the approach detailed in~\citet{yu2024mip}.
\begin{equation} \label{eq:cal_scale}
s_t = \dfrac{d_t}{f_i} + \varepsilon
\end{equation}
where $s_t$ represents the size of the corresponding Gaussian in world space, and $\varepsilon$ is a constant scalar. The scale of a camera view is set as the median value of the scales of all its sparse feature points.

\input{Figures/gaussian_init}

\subsubsection{Gaussian Densification} \label{sec:densification}
\label{sec:densification}
Previous works densify Gaussians mainly based on gradient magnitude~\cite{kerbl20233d} without explicitly targeting high-frequency regions with complex geometries and textures. In contrast, our approach leverages image-space information to directly initialize new Gaussians in the areas characterized by complex geometries, intricate textures, and high errors. An example of this initialization process is illustrated in Fig.~\ref{fig:ginit}. Specifically, we initialize new Gaussians from pixels in the following regions:

\textbf{Geometry/texture complex regions.} Tracking systems extract feature points from each frame to calculate the camera poses. These feature points are often located on high-contrast boundaries, which generally correspond to complex geometries or textures and require more Gaussians for accurate reconstruction. In contrast, textureless areas contain fewer feature points and consequently demand fewer Gaussians for reconstruction. Leveraging the feature point distribution from the tracking system provides a balanced approach to handling both high-frequency and low-frequency regions. 

\textbf{High error regions.} 
In addition to the tracked feature points, some regions with rich textures or complex geometries may still lack sufficient Gaussians. To address this, we calculate the SSIM (structural similarity index measure) between the rendered images $\bar{I}_i$ and the observed ground truth images $I_i$. Additional $k$ pixels for each keyframe are compensated in regions defined as $\bar{P}_i = \{\bar{p}_t\in \mathbb{R}^3 | SSIM(\bar{I}_i, I_i)[t] < \varepsilon_e\}$, where $\varepsilon_e$ is a predefined threshold. The depth values for these pixels are estimated using a pre-trained model~\cite{dexheimer2023depthcov}. Notably, since we only require a few sparse points to compensate for these regions, computing a complete depth map for all pixels is unnecessary. This approach is significantly faster and contrasts with other methods~\cite{sandstrom2024splat,zhang2024hi} that rely on pre-trained models to predict full-depth maps.

\subsubsection{Multi-level Occupancy Hash Voxel} \label{sec:MOHV}
Directly using all points in $P_i$ and $\bar{P}_i$ often leads to redundant and overlapping Gaussians, as many tracked points are clustered in similar positions. Additionally, scenes with varying levels of detail require dynamic adjustments: increasing Gaussian density in detailed regions, such as when zooming in, while avoiding excessive Gaussians in coarser areas. To eliminate redundancy and maintain efficiency, we propose a Multi-level Occupancy Hash Voxel (MOHV) structure to effectively remove redundant Gaussians and dynamically regulate their distribution in the world space across different levels.

$K$-nearest neighbors remove redundant Gaussians by rejecting those within a threshold distance of each other, but this approach becomes increasingly slow as the number of Gaussians grows in larger scenes. While occupancy voxels enable fast location queries to determine whether a position is occupied, their high memory requirements limit the resolution of fine-level voxels. To overcome these challenges, we adopt a multi-level hash structure inspired by Instant-NGP~\cite{muller2022instantngp}, which significantly reduces the memory consumption of occupancy voxels by leveraging the sparsity inherent in reconstructed scenes.
The MOHV module is defined by three parameters: number of levels $L$, initial scales $S_{\text{init}}$ for the coarsest level scales, and number of voxels per dimension $n$, resulting in a total of $n^3$ voxels for each level. 
The total memory cost of this structure is only $O(Ln^3)$. At the same time, it achieves a voxel resolution of $S_{\text{init}}/2^L$, as each successive level represents double the resolution of the previous one. 

\input{Figures/MOHV}

Fig.~\ref{fig:mohv} illustrates the high-level concept of the query and update operations in the MOHV module. When updating a position as occupied, the module marks all corresponding voxels up to the specified level, ensuring that finer-level occupancy remains unaffected. When querying a position, the module performs an AND operation on all occupancy information from the coarsest level to the specified level. This ensures that updates to finer occupancy levels are reflected in queries at coarser levels, aligning with our design objectives. Detailed algorithmic descriptions are provided in the supplementary.

Finally, MOHV removes Gaussians $\{P_i,\ \bar{P}_i\}$ located in occupied positions and updates its structure for the remaining Gaussians. The remaining Gaussians are initialized and added to the scene map with corresponding world positions, scales calculated through Eq.~\ref{eq:cal_scale}, and colors with corresponding pixels.

\subsection{Global Consistency Optimization} \label{sec:global_opt}
After distributing Gaussians in world space, their optimization is guided by selected views. Achieving a balance between local and global maps is challenging, as it requires both rapid convergence in newly observed regions and the preservation of previously reconstructed areas. To address this, the global consistency optimization module jointly optimizes Gaussians and online camera poses, ensuring a globally consistent reconstruction. This is accomplished through a carefully designed view selection strategy and optimization process, consisting of the following components:

\subsubsection{Keyframe Selection}
Using all frames for global optimization is redundant and usually yields lower quality. Instead, we select a subset of frames as keyframes for our global optimization. A frame $i$ is added as a keyframe if the overlap ratio $\text{covis}(i, j)$ between the current frame $i$ and the previous keyframe $j$ is below a threshold or if $t_k$ frames have elapsed since the last keyframe, accounting for textureless regions. The overlap ratio is calculated using tracked points instead of Gaussians to improve computational efficiency.

\input{Figures/optimize_windows}

\subsubsection{Optimization View Selection}
Unlike offline 3DGS optimization~\cite{kerbl20233d}, where all Gaussians are optimized simultaneously, online mapping requires incremental optimization. This makes it essential to balance newly observed views with historical ones. As shown in Fig.~\ref{fig:opt-windows}, we propose a local and global camera selection strategy to achieve fast convergence for new frames while maintaining global consistency in previously reconstructed regions. Given the current view camera $c_i$, the local and global cameras are defined as:

\textbf{Local Cameras.} Local cameras aim to optimize newly observed regions with multi-view constraints. In our experiments, we set the number of additional local views, $n_{\text{local}}$, to $1$, in addition to the current view. To ensure sufficient overlap with the current view, we maintain a local bank of size $\bar{n}_{\text{local}}$. A new frame is added to the bank every $t_{\text{local}}$ frames and the oldest frame is discarded once the bank exceeds its maximum size. This bank selects the $n_{\text{local}}$ views with the largest overlap with the current view for multi-view joint optimization.

\textbf{Global Cameras.} 
Local views converge quickly but cannot ensure a globally consistent map. Relying solely on local views often results in overfitting to specific regions, leading to poor global maps due to forgetting issues and camera drift. To address this, we also select $n_{\text{global}}$ views from historical keyframes. However, randomly sampling from all historical keyframes creates an imbalance, with earlier frames being selected more frequently than recent ones. To mitigate this issue, we sample the historical keyframes based on the following probability:
\begin{equation}
    \text{prob}_i[j] = \text{normalize}(e^{\sigma_1\cdot (j-i)}\cdot e^{\sigma_2 \cdot \textrm{err}(j)}),
\end{equation}
where $\textrm{err}(j)$ refers to the Mean Absolute Error of frame $j$, updated every time frame $j$ is optimized. The first term in the probability distribution adjusts the selection to prioritize newly added keyframes, while the second term emphasizes under-optimized keyframes.

\subsubsection{Camera Refinement}
Although the online tracking system provides reasonably accurate camera poses, they are insufficient for reconstructing high-quality maps. To enhance reconstruction quality, we further optimize the poses of keyframes during the mapping process.

The scene's Gaussian map is optimized for each keyframe using rendering losses, incorporating the current view along with the selected local and global views at every step. Further details on the optimization parameters are provided in the supplementary.

%% file: Figures/gaussian_init.tex
\begin{figure}
    \centering
    \begin{overpic}[width=\columnwidth]{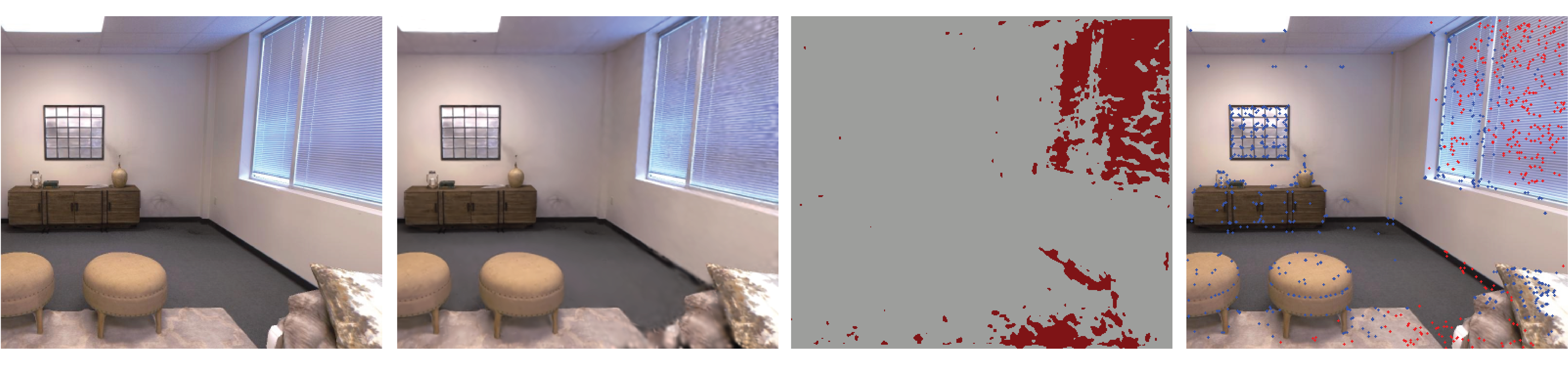}
    \put(6, -0.2){\color{black}\footnotesize{Reference}}
    \put(27.5, -0.2){\color{black}\footnotesize{Online Rendered}}
    \put(57, -0.2){\color{black}\footnotesize{Error map}}
    \put(77.5, -0.2){\color{black}\footnotesize{Initialized points}}
    \end{overpic}
    \caption{A visualization of pixels used for initialization. In the error map, red regions indicate high-error areas. In the right-most image, {\color{red}red} points represent pixels used for error compensation, while {\color{blue}blue} points correspond to pixels in geometry- or texture-complex regions.
    }
    \label{fig:ginit}
\end{figure}

%% file: Figures/MOHV.tex
\begin{figure}
    \centering
    \begin{overpic}[width=\columnwidth]{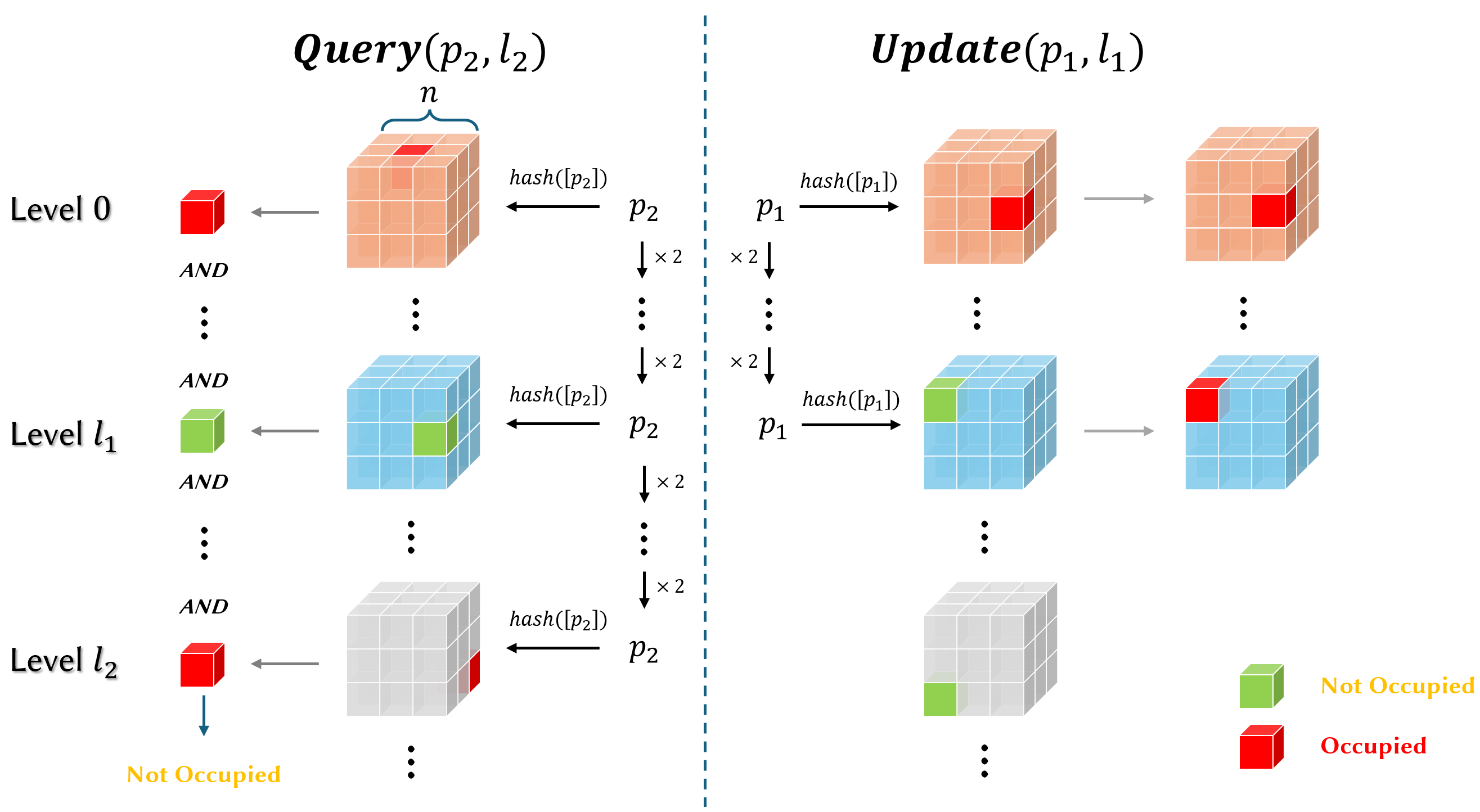}
    \end{overpic}
    \caption{The high-level concept of MOHV. It updates and queries multi-level voxels up to a given level $l$ to maintain Gaussian distributions for capturing details at various levels.
    }
    \label{fig:mohv}
\end{figure}

%% file: Figures/optimize_windows.tex
\begin{figure}
    \centering
    \begin{overpic}[width=\columnwidth]{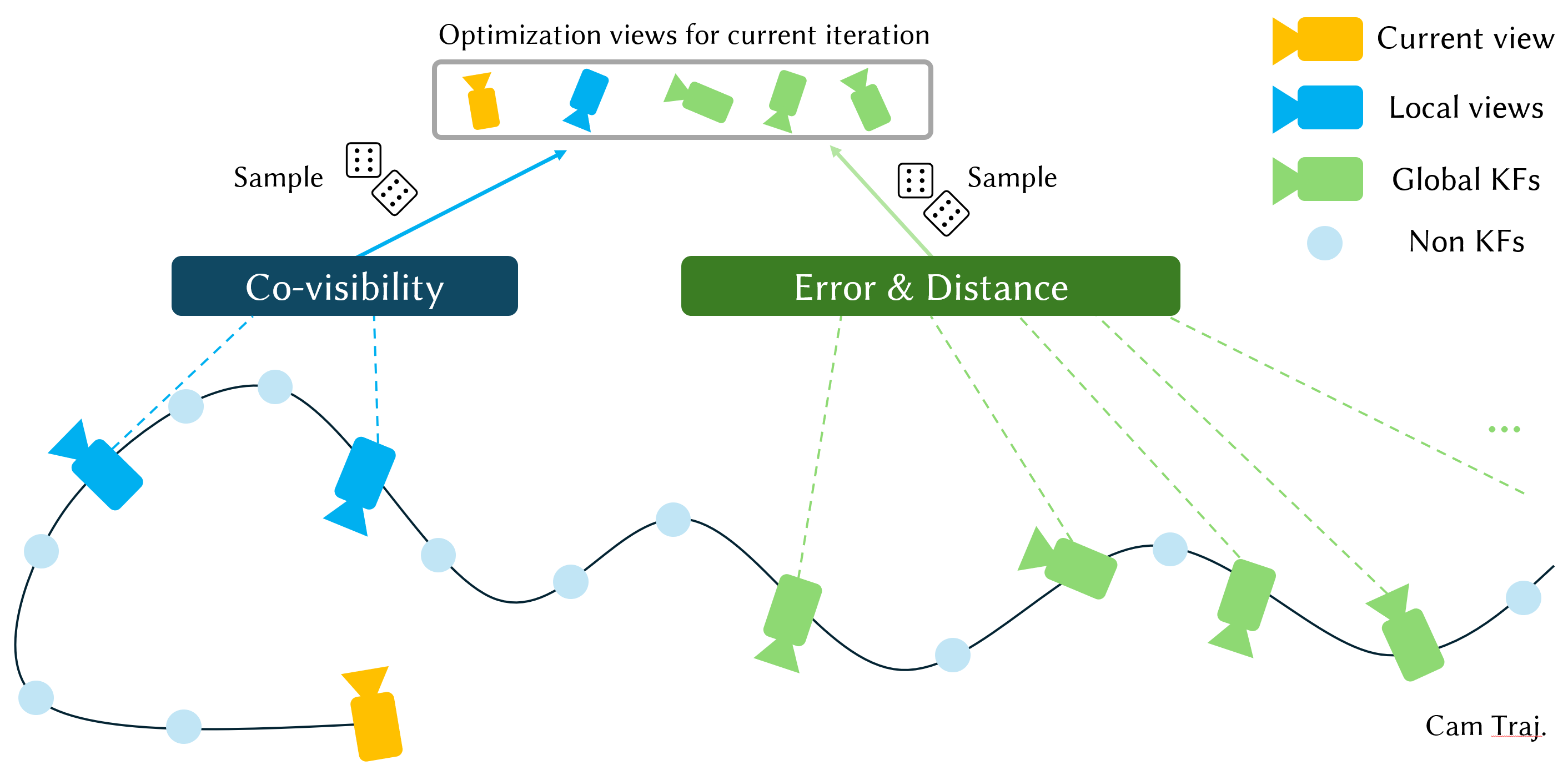}
    \end{overpic}
    \caption{High-level concept of optimization view selection. Our method samples local views based on covisibility, and global views based on their L1 error values and distance to the current view.
    }
    \label{fig:opt-windows}
\end{figure}

%% file: experiments.tex
\section{Experiments}
\subsection{Dataset}
To demonstrate the robustness of our pipeline, we evaluate our method on three datasets: TUM-RGBD~\cite{sturm2012benchmark}, Replica \cite{straub2019replica}, as well as our own sequences captured using Aria glasses~\cite{engel2023project}. The TUM-RGBD dataset consists of real-world RGB-D images but includes challenging sequences with severe motion blur, which often degrade reconstruction quality. The Replica dataset provides highly accurate depth maps since it is re-rendered from reconstructed 3D models.
Additionally, we use Aria glasses to capture indoor and outdoor sequences with varying geometric complexity. All frames captured with Aria glasses undergo consistent pre-processing operations before being tested on our pipeline and baselines, ensuring a fair comparison.

\subsection{Baselines}
We compare our method against existing dense SLAM baselines, including both monocular and RGB-D approaches. Specifically, we compare with prior works~\cite{Matsuki:Murai:etal:CVPR2024, huang2024photo, sandstrom2024splat} for monocular SLAM baselines and~\cite{Matsuki:Murai:etal:CVPR2024, peng2024rtg, ha2025rgbd} for RGB-D SLAM baselines. We also compare with concurrent work~\cite{zhang2024hi} using their released codes.
For our custom-captured sequences, where depth maps are unavailable, we compare with monocular baselines only. To ensure fairness, all evaluations are performed on the full sequence of images at their original resolution.

Note that several methods (MonoGS~\cite{Matsuki:Murai:etal:CVPR2024}, Splat-SLAM~\cite{sandstrom2024splat}, Hi-SLAM2~\cite{zhang2024hi}) include a post-refinement step unrelated to the online reconstruction process. To ensure a fair comparison, we evaluate the results both before and after the post-refinement step separately. For Aria sequences, we run $26K$ steps for baselines in the post-refinement, while our method uses only $1K$ post-refinement steps for all scenes since it is not an essential step for ours.

\subsection{Mapping and Rendering Quality}

\input{Tables/comparison_rendering}

Our main objective is to reconstruct high-quality maps with a photorealistic appearance. Table~\ref{tlb:comparison_render} presents a quantitative comparison of our pipeline against existing monocular and RGB-D baselines. We use peak signal-to-noise ratio (PSNR), structural similarity index (SSIM) \cite{wang2004image}, and perceptual similarity (LPIPS)~\cite{zhang2018unreasonable} as our evaluation metrics. PSNR and SSIM measure overall reconstruction quality, while LPIPS emphasizes the preservation of fine details.

Our method outperforms all monocular baselines and surpasses RGB-D baselines in real-world captured scenarios. While RGB-D methods excel on the synthetic Replica dataset, which provides perfect depth maps, their quality degrades in real-world cases with imperfect depth data. The Aria dataset, specifically captured to test scenes with varying levels of detail, highlights our method's superior performance, particularly in LPIPS, demonstrating enhanced detail preservation compared to other baselines. 

Qualitative comparisons are shown in Fig.~\ref{fig:aria_comparison}, Fig.~\ref{fig:replica_comparison}, and Fig.~\ref{fig:tum_comparison}. In complex and challenging scenes with varying detail levels, our pipeline reconstructs significantly better details, while other baselines produce overly blurred results, even after applying a global post-refinement process.

\subsection{Tracking Accuracy}
\input{Tables/tracking}
Although our framework is designed to be compatible with various tracking systems and does not specifically focus on tracking accuracy, we analyze the tracking performance to better understand its impact on the quality of our online reconstruction framework. Table~\ref{tlb:tracking} presents the tracking errors of different methods. While our framework exhibits slightly lower tracking accuracy, it consistently delivers higher-quality reconstruction maps, demonstrating the effectiveness of our online reconstruction pipeline. Improving integration with tracking systems is left for future work to further enhance reconstruction quality.

\subsection{Speed Performance}
\input{Tables/performance}
All experiments were conducted on a Fedora machine with an AMD Ryzen Threadripper PRO 3975WX and an NVIDIA RTX 4090. As shown in Table~\ref{tlb:speed}, our method achieves approximately 10 FPS on small-scale scenes (TUM and Replica) and around 5 FPS on relatively larger scenes (Aria) while maintaining high reconstruction quality. Note that Photo-SLAM~\cite{huang2024photo}, implemented entirely in C++, is several orders of magnitude faster than other pipelines built with Python. 

A detailed time breakdown is provided in Table~\ref{tlb:speed_breakdown}, demonstrating that our global camera selection, depth estimation, and MOHV modules are sufficiently efficient and do not constitute the pipeline's bottleneck. Additional details can be found in the supplementary material.

\subsection{Ablation Studies}
\input{Tables/ablation_studies_gs}
In this section, we perform comprehensive ablation studies to evaluate the effectiveness of individual modules in our framework. As our designs primarily focus on reconstructing details across various levels of the scene, the ablation studies are conducted on the Aria sequences, which feature scenes with diverse geometric complexity. 

\input{Figures/abl_err_comp}

\textbf{Error region compensation.} Tracking systems without dense depth maps typically provide only sparse 3D tracked points, which are insufficient for reconstructing fine details. Our error region compensation module adds supplementary points in high-error and under-optimized regions. As illustrated in Fig.~\ref{fig:abl_err_comp}, incorporating error region compensation enhances detail in areas where tracked points are sparse or missing. Without it, the textures on the bookshelf and the details on the plant are noticeably missing in Fig.~\ref{fig:abl_err_comp}.
Note that while increasing the number of tracked points by adjusting the tracking system's threshold is possible, doing so may adversely affect the overall tracking quality.

\textbf{Multi-level occupancy hash voxel.} The MOHV module removes redundant Gaussians within local regions based on camera scales. As shown in Table.~\ref{tlb:abation_studies_gs} and Fig.~\ref{fig:abl_err_comp}, although with about $100$K more Gaussians, the final reconstructed quality does not noticeably improve. This demonstrates that MOHV preserves reconstruction quality while reducing the number of Gaussians by approximately $30$\%. By effectively controlling the growth of Gaussians, this module enables scalability to large-scale scenes without redundant overhead.

\input{Tables/ablation_studies_opt}
\textbf{Global view optimization.} Global view optimization is essential for maintaining a globally consistent map, preventing forgetting issues, and balancing newly observed regions with previously reconstructed ones. For fairness, we add the same number of local views to ensure an equal optimization budget when global view optimization is removed. As shown in Table~\ref{tlb:abation_studies_opt} and Fig.~\ref{fig:abl_global_cam}, the quality of previously reconstructed regions significantly deteriorates when the optimization focuses solely on newly observed regions.

\textbf{Camera refinement.} Camera refinement is used to globally optimize both the Gaussians and keyframes camera poses to improve reconstruction quality. As shown in Fig.~\ref{fig:abl_cam_refine}, joint camera refinement improves the details in the reconstruction.

\subsection{Alternative Tracking Systems} \label{sec:aria_tracking}
\input{Tables/aria_tracking}
Our framework is a general online mapping system, which is not limited to the ORB-SLAM3 tracking system. In this section, we also show the results of using the Aria tracking system~\cite{engel2023project} which uses Aria's two SLAM cameras and an inertial measurement unit (IMU) to perform stereo tracking. Notably, our mapping system does not directly use SLAM cameras and IMU's information and only takes online tracked poses and sparse points as inputs to align our monocular online reconstruction settings. Table.~\ref{tlb:aria_tracking} and Fig.~\ref{fig:diff-track_comparison} compare our method using ORB-SLAM3 and Aria tracking systems. The results demonstrate comparable performance across both systems. The slightly lower PSNR/SSIM scores with Aria tracking can be attributed to its sparse feature points, which prioritize mid-range objects over distant ones to enhance detail reconstruction. This experiment highlights the robustness and versatility of our framework, showcasing its compatibility with different tracking systems.

\subsection{Post Refinement}
Our method is designed for high-quality online reconstruction, with the option of a post-refinement process to further enhance reconstruction quality starting from the results of our online pipeline. Table.~\ref{tlb:aria_tracking} and Fig.~\ref{fig:diff-track_comparison} show our method with different steps in the post-refinement process as well as the offline 3DGS baseline~\cite{kerbl20233d}. Although offline 3DGS is a global optimization approach, our methods with more post-refinement steps show better reconstruction results while requiring less time and fewer Gaussians. 

%% file: Tables/comparison_rendering.tex
\begin{table}
    \begin{center}

    \renewcommand{\arraystretch}{1.0} {
    \caption{Comparison of reconstruction rendering quality on different datasets. \textbf{Bold} refers to the best across all categories and \colorbox{green!20}{green} refers to the best of each category. Numbers of RTG-SLAM (Replica), GS-ICP (TUM and Replica) and Photo-SLAM (TUM and Replica) are taken from their original papers. Other numbers are calculated through their released codes. }
        \begin{tabular}{lll|w{c}{4em}w{c}{4em}w{c}{4em}}
        \toprule

        & Methods & Metrics & TUM & Replica & Aria \\ \hline

        \multirow{9}{1em}{\rotatebox[origin=c]{90}{RGB-D}}   & \multirow{3}{2em}{MonoGS} & PSNR & 17.90 & 36.67 & n/a \\
                                                                     & & SSIM & 0.716 & 0.958 & n/a  \\
                                                                     & & LPIPS & 0.322 & 0.072 & n/a  \\ \cline{2-6}
                                                                     & \multirow{3}{1em}{RTG-SLAM} & PSNR & 19.44 & 35.43 & n/a \\
                                                                     & & SSIM & 0.760 & 0.982 & n/a \\
                                                                     & & LPIPS & 0.408 & 0.109 & n/a \\ \cline{2-6}
                                                                     & \multirow{3}{1em}{GS-ICP} & PSNR & \colorbox{green!20}{20.72} & \textbf{\colorbox{green!20}{38.83}} & n/a \\
                                                                     & & SSIM & \colorbox{green!20}{0.768} & \textbf{\colorbox{green!20}{0.975}} & n/a  \\
                                                                     & & LPIPS & \colorbox{green!20}{0.218} & \textbf{\colorbox{green!20}{0.041}} & n/a \\ \hline
        \multirow{15}{1em}{\rotatebox[origin=c]{90}{Monocular}}    & \multirow{3}{1em}{MonoGS} & PSNR & 17.54 & 27.38 & 18.34  \\ 
                                                                     & & SSIM & 0.698 & 0.860 & 0.475 \\
                                                                     & & LPIPS & 0.341 & 0.261 & 0.700  \\ \cline{2-6}
                                                                     & \multirow{3}{1em}{Photo-SLAM} & PSNR & 20.54 & 33.30 & 23.40  \\
                                                                     & & SSIM & 0.720 & 0.926 & 0.615 \\
                                                                     & & LPIPS & 0.211 & 0.078 & 0.477 \\ \cline{2-6}
                                                                     & \multirow{3}{1em}{Splat-SLAM} & PSNR & 22.34 & 30.37 & 21.58 \\
                                                                     & & SSIM & 0.731 & 0.886 & 0.561  \\
                                                                     & & LPIPS & 0.353 & 0.221 & 0.606 \\ \cline{2-6}
                                                                     & \multirow{3}{1em}{Hi-SLAM2} & PSNR & 20.09 & 30.74 & 19.60 \\
                                                                     &  & SSIM & 0.680 & 0.897 & 0.520 \\
                                                                     &  & LPIPS & 0.379 & 0.208 & 0.675 \\ \cline{2-6}
                                                                     & \multirow{3}{1em}{\textbf{Ours}} & PSNR & \colorbox{green!20}{25.45} & \colorbox{green!20}{35.85} &  \colorbox{green!20}{26.15} \\
                                                                     & & SSIM & \colorbox{green!20}{0.866} & \colorbox{green!20}{0.956} & \colorbox{green!20}{0.678} \\
                                                                     & & LPIPS & \colorbox{green!20}{0.165} & \colorbox{green!20}{0.071} & \colorbox{green!20}{0.338} \\ \hline 
        \multirow{12}{1em}{\rotatebox[origin=c]{90}{Monocular (w Post Refinement)}}    & \multirow{3}{1em}{MonoGS} & PSNR & 22.04 & 30.13 & 21.05  \\ 
                                                                     & & SSIM & 0.737 & 0.900 & 0.555 \\
                                                                     & & LPIPS & 0.326 & 0.193 & 0.662  \\ \cline{2-6}
                                                                     & \multirow{3}{1em}{Splat-SLAM} & PSNR & 25.53 & 33.72 & 24.44 \\
                                                                     & & SSIM & 0.801 & 0.938 & 0.618  \\
                                                                     & & LPIPS & 0.251 & 0.117 & 0.490 \\ \cline{2-6}
                                                                     & \multirow{3}{1em}{Hi-SLAM2} & PSNR & 23.52 & 36.69 & 25.64 \\
                                                                     & & SSIM & 0.805 & 0.953 & 0.665 \\
                                                                     & & LPIPS & 0.242 & 0.113 & 0.414 \\ \cline{2-6}
                                                                     & \multirow{3}{1em}{\textbf{Ours}} & PSNR & \textbf{\colorbox{green!20}{26.18}} & \colorbox{green!20}{36.89} & \textbf{\colorbox{green!20}{26.62}}  \\
                                                                     &  & SSIM & \textbf{\colorbox{green!20}{0.874}} & \colorbox{green!20}{0.962} & \textbf{\colorbox{green!20}{0.693}} \\
                                                                     &  & LPIPS & \textbf{\colorbox{green!20}{0.154}} & \colorbox{green!20}{0.061} & \textbf{\colorbox{green!20}{0.324}} \\ 
        
        \bottomrule
        \end{tabular}
        \label{tlb:comparison_render}
    }
    \end{center}
\end{table}

%% file: Tables/tracking.tex
\begin{table}
    \begin{center}

    \setlength{\tabcolsep}{2.2mm}{
    \renewcommand{\arraystretch}{1.0} {
    \caption{Tracking accuracy of different monocular baselines and our methods. Numbers represent absolute trajectory error (ATE) root mean square error (RMSE) in cm. All baseline numbers are taken from their original paper except Hi-SLAM2 (TUM) and MonoGS (Replica) since they didn't include those trajectory accuracy in their paper.  }
        \begin{tabular}{l|ccc}
        \toprule
                 & MonoGS & PhotoSLAM & SplatSLAM \\ 
            TUM  & 3.96 &  1.26 &  \textbf{1.1} \\ 
            Replica & 22.03 & 1.09 & 0.35  \\ \hline
                 & Hi-SLAM2   & ORB-SLAM3 & Ours  \\ 
            TUM & 1.32 & 1.98 & 1.87  \\
            Replica & \textbf{0.26} & 3.88 & 3.75 \\
        \bottomrule
        \end{tabular}
        \label{tlb:tracking}
    }
        
    }
    \end{center}
\end{table}

%% file: Tables/performance.tex
\begin{table}
    \begin{center}

    \setlength{\tabcolsep}{5mm}{
    \renewcommand{\arraystretch}{1.05} {
    \caption{Speed performance of different methods. Numbers represent frames per second (FPS).}
        \begin{tabular}{l|ccc}
        \toprule

                          & TUM    & Replica     &  Aria \\ \hline
            MonoGS        &  2.91  &    1.43     &  1.85    \\
            Photo-SLAM    &  65.87 &    41.65    &  10.02 \\
            Splat-SLAM    &  3.62  &    1.01     &  0.46  \\
            Hi-SLAM2      &  13.99 &    15.08    &  3.35  \\
            Ours          &  11.28 &    9.34    &  4.55  \\
        
        \bottomrule
        \end{tabular}
        \label{tlb:speed}
    }
        
    }
    \end{center}
\end{table}

\begin{table}
    \begin{center}

    \setlength{\tabcolsep}{2mm}{
    \renewcommand{\arraystretch}{1.05} {
    \caption{Time breakdown of each component. The reported numbers represent the percentage of time consumed by each component relative to the overall process. "Others" primarily includes data transfer and preprocessing operations. }
        \begin{tabular}{l|cccccc}
        \toprule

                          & Optim  & Cam Select  & Depth Est. & MOHV & Others  \\ \hline
            PCT (\%)        &  91.44  & 0.79   & 1.63     &  0.87  & 5.27  \\
        
        \bottomrule
        \end{tabular}
        \label{tlb:speed_breakdown}
    }
        
    }
    \end{center}
\end{table}

%% file: Tables/ablation_studies_gs.tex
\begin{table}
    \begin{center}

    \setlength{\tabcolsep}{2.5mm}{
    \renewcommand{\arraystretch}{1.05} {
    \caption{Ablation studies of our Gaussian management module. Numbers are averaged from all Aria sequences. \textbf{Bold} refers to the best and \underline{underline} refers to the second best.}
        \begin{tabular}{l|ccccc}
        \toprule

                       & PSNR & SSIM & LPIPS & \# Gaussians  \\ \hline
        w/o EC         & 24.87 &   0.647   & 0.407      &   123,396        \\
        w/o MOHV       &  \underline{26.04}    & \textbf{0.680}     & \textbf{0.333}       & 445,764          \\ 
        
        w/o EC \& MOHV & 25.30 &   0.661   & 0.375      &   221,370        \\ \hline
        Ours full      &  \textbf{26.15}    & \underline{0.678}      & \underline{0.338}       & 340,962          \\

        \bottomrule
        \end{tabular}
        \label{tlb:abation_studies_gs}
    }
        
    }
    \end{center}
\end{table}

%% file: Figures/abl_err_comp.tex
\begin{figure}
    \centering
    \begin{overpic}[width=\columnwidth]{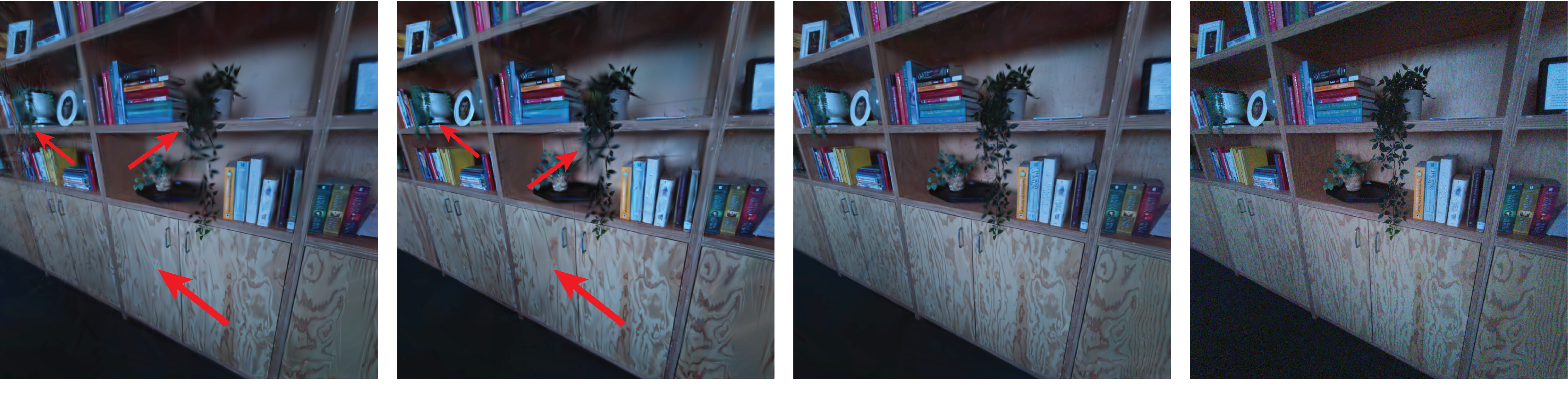}
    \put(6, -0.9){\color{black}\footnotesize{w/o EC}}
    \put(27, -0.9){\color{black}\footnotesize{w/o EC \& MOHV}}
    \put(56.5, -0.9){\color{black}\footnotesize{Ours full}}
    \put(81, -0.9){\color{black}\footnotesize{Reference}}
    \end{overpic}
    \caption{Ablation studies of Gaussian management module. MOHV refers to Multi-level Occupancy Hash Voxels and EC refers to Error Compensation. Zoom in for details.}
    \label{fig:abl_err_comp}
\end{figure}
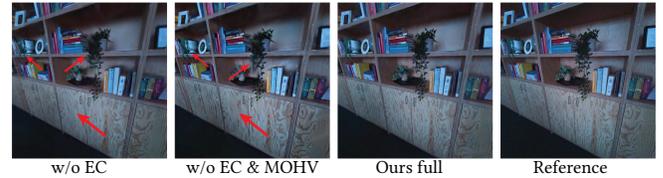

%% file: Tables/ablation_studies_opt.tex
\begin{table}
    \begin{center}

    \setlength{\tabcolsep}{3.5mm}{
    \renewcommand{\arraystretch}{1.05} {
    \caption{Ablation studies of our global consistent optimization module. \textbf{Bold} refers to the best.}
        \begin{tabular}{l|ccc}
        \toprule

                       & PSNR & SSIM & LPIPS  \\ \hline
        w/o global cams  & 17.68 &   0.460   & 0.595             \\
        w/o cam refinement      &  25.52    & 0.666      & 0.341                 \\ \hline
        Ours full     &  \textbf{26.15}    & \textbf{0.678}      & \textbf{0.338}            \\

        \bottomrule
        \end{tabular}
        \label{tlb:abation_studies_opt}
    }
        
    }
    \end{center}
\end{table}

%% file: Tables/aria_tracking.tex
\begin{table}
    \begin{center}

    \footnotesize
    \renewcommand{\arraystretch}{1.2} {
    \caption{Comparison of rendering quality on reconstructed maps with different tracking systems and offline 3DGS on Aria sequences. PR. refers to the post-refinement process. PR. steps in offline 3DGS refers to the total optimization steps.}
        \begin{tabular}{l|ccccccc}
        \toprule

                & PSNR & SSIM & LPIPS & PR. Steps & \# Gs & Time \\ \hline
        Ours & 26.15  & 0.678 & 0.338 & 0 & 341K & 4m1s  \\
        Ours (w PR.) & 26.62  & 0.693 & 0.324 & 1K & 341K & 4m56s  \\
        Ours (w PR.) & \textbf{27.44}  & \textbf{0.711} & 0.297 & 20K & 341K & 9m32s  \\
        Ours (Aria) & 25.07 & 0.659 & 0.330 & 0 & 700K & 3m55s \\
        Ours (Aria, w PR) & 25.67 & 0.675 & 0.315 & 1K & 700K & 4m32s \\
        Ours (Aria, w PR) & 25.99 & 0.682 & \textbf{0.294} & 20K & 700K & 10m24s \\
        Offline 3DGS & 26.60 & 0.696 & 0.305 & 150K & 1,781K & 1h6m21s \\

        \bottomrule
        \end{tabular}
        \label{tlb:aria_tracking}
    }
    \end{center}
\end{table}

%% file: conclusion.tex
\section{Conclusion}
In this work, we present a high-quality online reconstruction pipeline for reconstructing environments from monocular inputs. Our pipeline incorporates a hierarchical Gaussian management module and a global consistency optimization module, enabling the maintenance of Gaussians to capture details across various levels while remaining computationally efficient.

However, our method has certain limitations. One notable limitation arises when the tracking system loses tracking or when trajectory accumulation errors become significant. In the future, our method could be improved by explicitly addressing significant camera shifting issues through loop closure, extending the pipeline's applicability to even larger scenes.

%% file: Figures/Aria_comparison.tex
\begin{figure*}
    \centering
    \begin{overpic}[width=\linewidth]{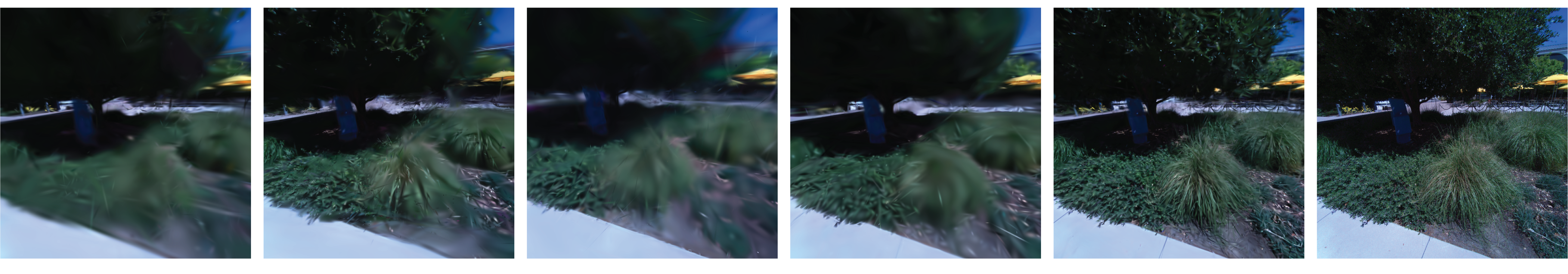}
    \end{overpic}
    \begin{overpic}[width=\linewidth]{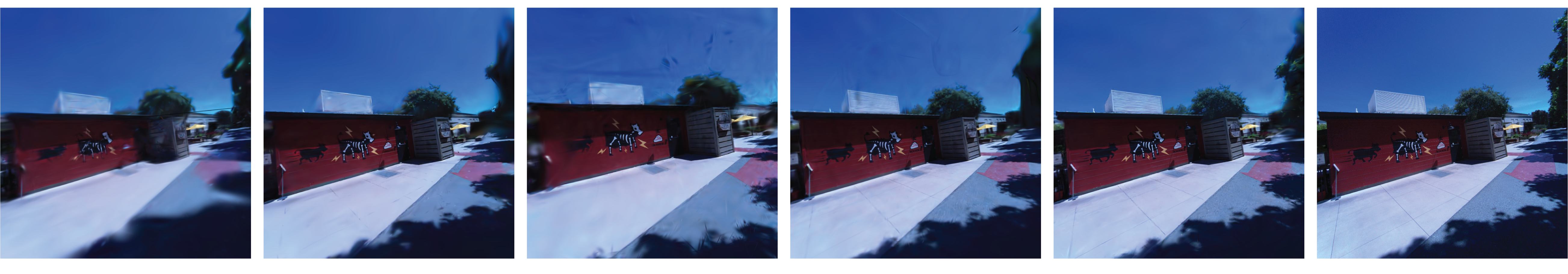}
    \end{overpic}
    \begin{overpic}[width=\linewidth]{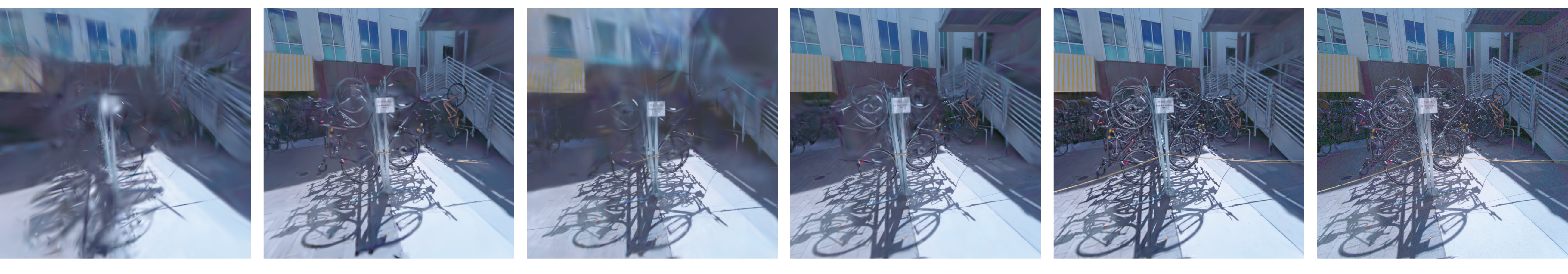}
    \end{overpic}
    \begin{overpic}[width=\linewidth]{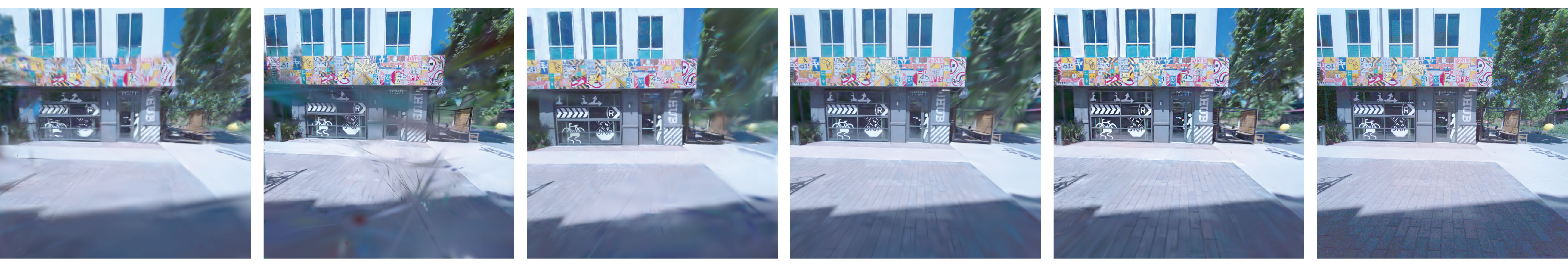}
    \end{overpic}
    \begin{overpic}[width=\linewidth]{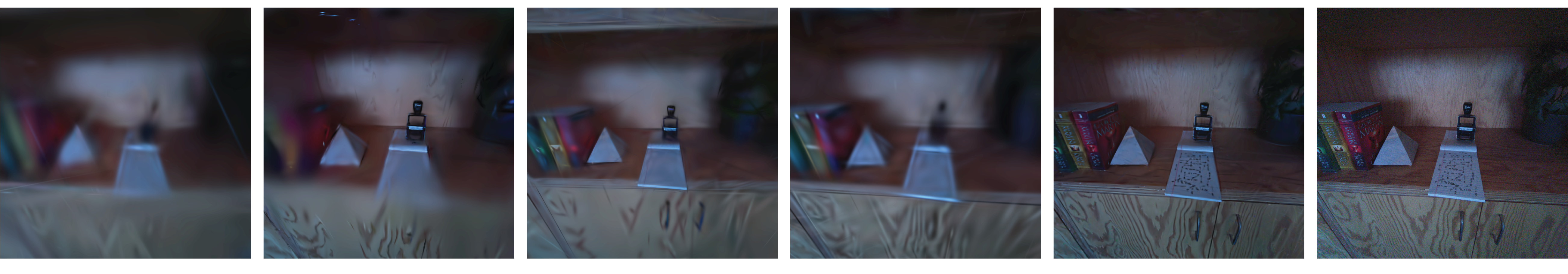}
    \end{overpic}
    \begin{overpic}[width=\linewidth]{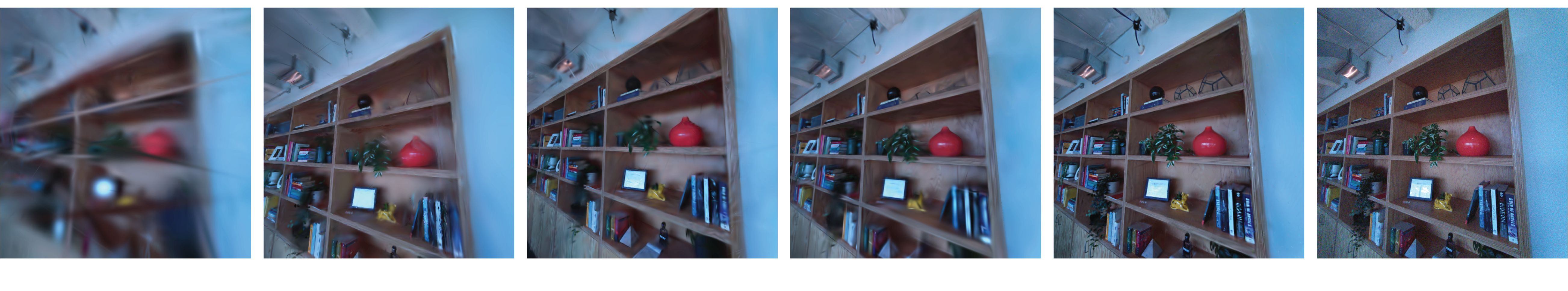}
    \put(5.75, 0.4){\color{black}\small{MonoGS}}
    \put(2.5, -1.4){\color{black}\small{(w Post-refinement)}}
    \put(21, 0.4){\color{black}\small{PhotoSLAM}}
    \put(38, 0.4){\color{black}\small{SplatSLAM}}
    \put(34.75, -1.4){\color{black}\small{(w Post-refinement)}}
    \put(55.5, 0.4){\color{black}\small{Hi-SLAM2}}
    \put(52.25, -1.4){\color{black}\small{(w Post-refinement)}}
    \put(73, 0.4){\color{black}\small{Ours}}
    \put(88, 0.4){\color{black}\small{Reference}}
    \end{overpic}
    \caption{Qualitative comparison on Aria captured sequences. Our method captures finer details, while other baselines produce over-blurred results. Notably, even after applying $26$K post-refinement steps for the baselines, they still exhibit poorer details compared to our method. This demonstrates that high-quality, fine-level details cannot be achieved solely by increasing the number of optimization iterations. Zoom in for more details.}
    \label{fig:aria_comparison}
\end{figure*} 

%% file: Figures/image_only.tex
\begin{figure*}
    \centering
    \begin{overpic}[width=\linewidth]{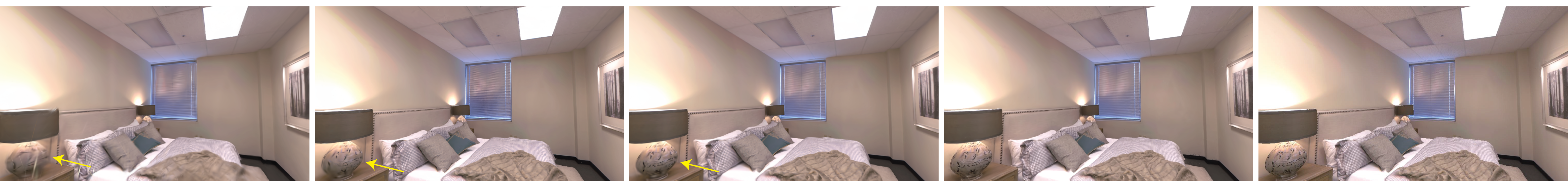}
    \end{overpic}
    \begin{overpic}[width=\linewidth]{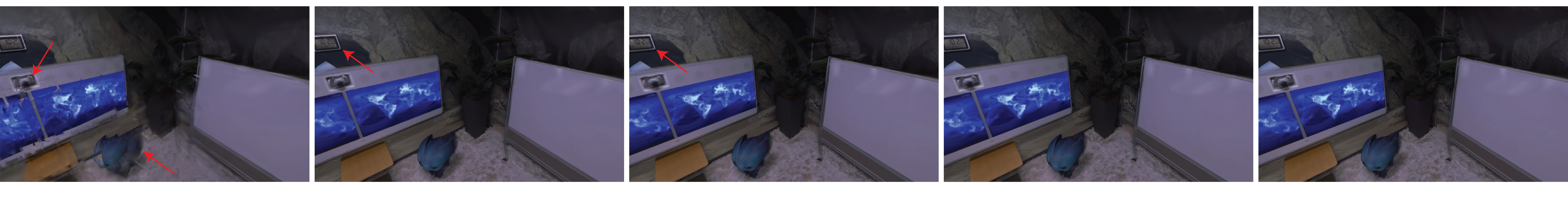}
    \put(7.75, 0.4){\color{black}\small{MonoGS}}
    \put(4.5, -1.4){\color{black}\small{(w Post-refinement)}}
    \put(27, 0.4){\color{black}\small{SplatSLAM}}
    \put(23.75, -1.4){\color{black}\small{(w Post-refinement)}}
    \put(46.5, 0.4){\color{black}\small{Hi-SLAM2}}
    \put(43.25, -1.4){\color{black}\small{(w Post-refinement)}}
    \put(68, -0.4){\color{black}\small{Ours}}
    \put(87, -0.4){\color{black}\small{Reference}}
    \end{overpic}
    \caption{Qualitative comparison on Replica dataset. }
    \label{fig:replica_comparison}
\end{figure*} 

\begin{figure*}
    \centering
    \begin{overpic}[width=\linewidth]{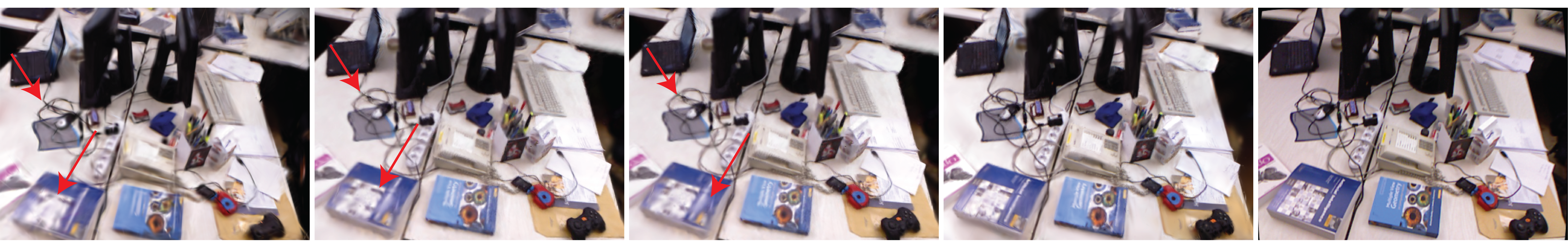}
    \end{overpic}
    \begin{overpic}[width=\linewidth]{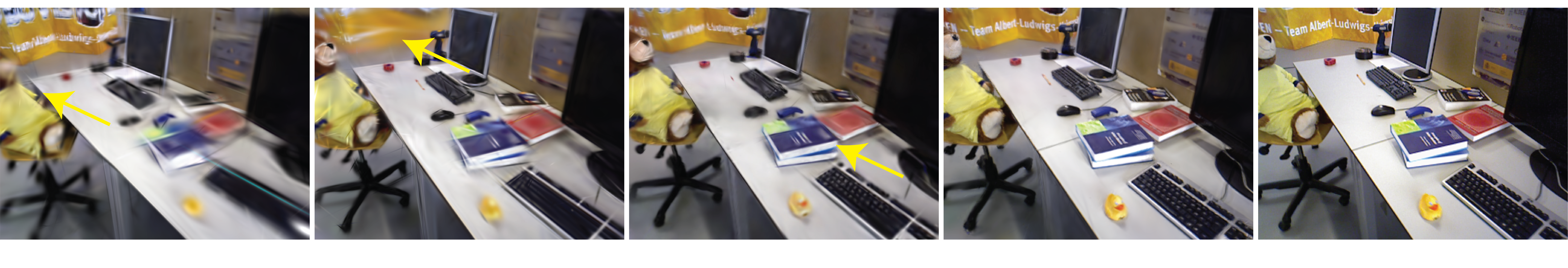}
    \put(7.75, 0.4){\color{black}\small{MonoGS}}
    \put(4.5, -1.4){\color{black}\small{(w Post-refinement)}}
    \put(27, 0.4){\color{black}\small{SplatSLAM}}
    \put(23.75, -1.4){\color{black}\small{(w Post-refinement)}}
    \put(46.5, 0.4){\color{black}\small{Hi-SLAM2}}
    \put(43.25, -1.4){\color{black}\small{(w Post-refinement)}}
    \put(68, -0.4){\color{black}\small{Ours}}
    \put(87, -0.4){\color{black}\small{Reference}}
    \end{overpic}
    \caption{Qualitative comparison on TUM dataset. }
    \label{fig:tum_comparison}
\end{figure*} 

\begin{figure*}
    \centering
    \begin{overpic}[width=\linewidth]{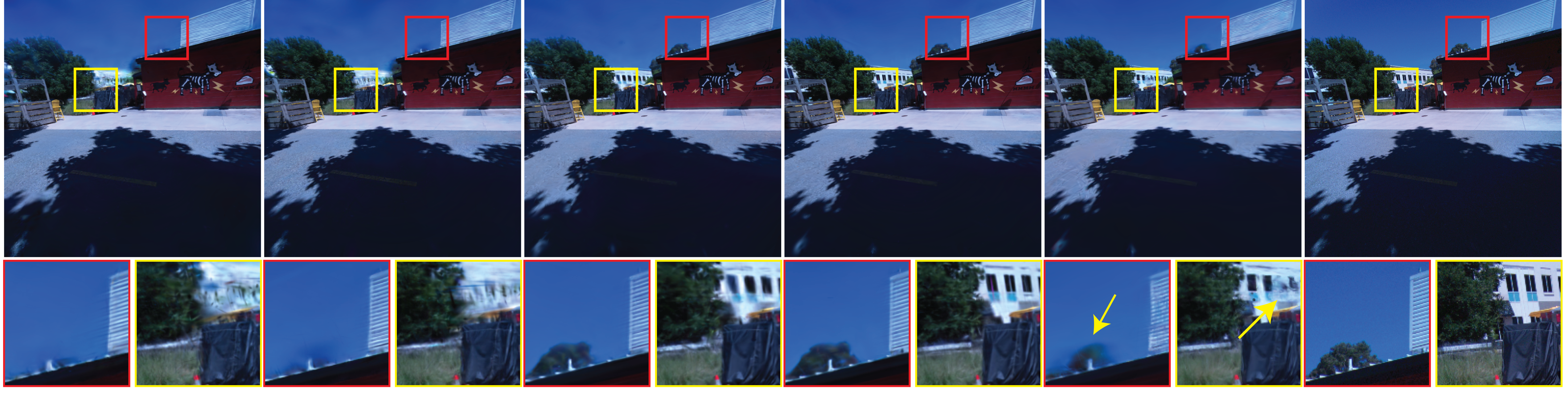}
    \put(5.75, 0.4){\color{black}\small{Ours-Aria}}
    \put(21.5, 0.4){\color{black}\small{Ours-Aria}}
    \put(19, -1.4){\color{black}\small{(w 20K PR. steps)}}
    \put(40, 0.4){\color{black}\small{Ours}}
    \put(56.5, 0.4){\color{black}\small{Ours}}
    \put(53, -1.4){\color{black}\small{(w 20K PR. steps)}}
    \put(73, 0.4){\color{black}\small{3DGS}}
    \put(88, 0.4){\color{black}\small{Reference}}
    \end{overpic}
    \caption{Qualitative comparison of our methods with different tracking systems and offline 3DGS method. PR. refers to the post-refinement process. }
    \label{fig:diff-track_comparison}
\end{figure*}

\begin{figure*}[ht]
    \flushleft
    \begin{minipage}{\columnwidth}
    \begin{overpic}[width=\columnwidth]{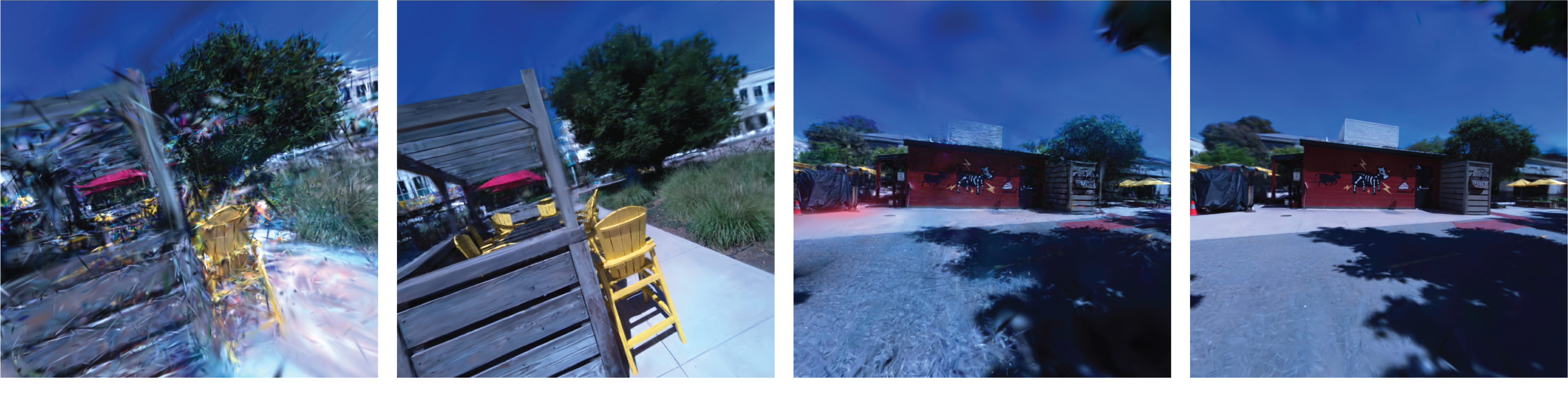}
    \put(6, -0.6){\color{black}\footnotesize{w/o global}}
    \put(32, -0.6){\color{black}\footnotesize{w global}}
    \put(56.5, -0.6){\color{black}\footnotesize{w/o global}}
    \put(82.5, -0.6){\color{black}\footnotesize{w global}}
    \put(0.75, 23){\color{white}\footnotesize{Frame 475}}
    \put(51.5, 23){\color{white}\footnotesize{Frame 950}}
    \end{overpic}
    \caption{Ablation studies of global view optimization. 
    }
    \label{fig:abl_global_cam}
    \end{minipage}
    \hfill
    \begin{minipage}{\columnwidth}
    \begin{overpic}[width=\columnwidth]{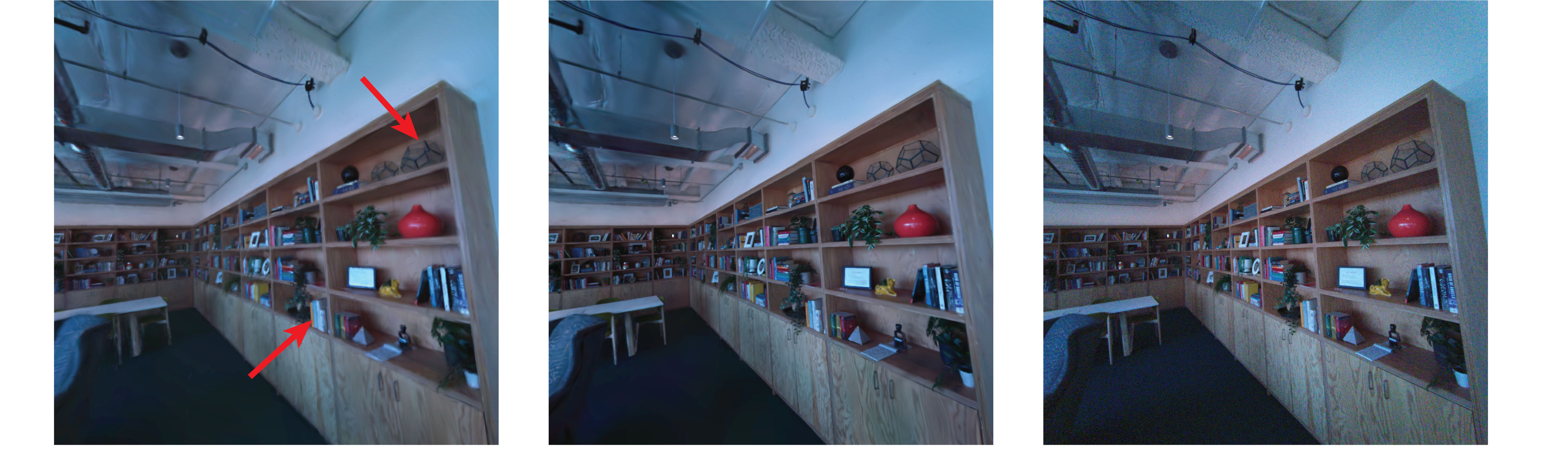}
    \put(5, -0.7){\color{black}\footnotesize{w/o cam refinement}}
    \put(38, -0.7){\color{black}\footnotesize{w cam refinement}}
    \put(76, -0.7){\color{black}\footnotesize{Reference}}
    \end{overpic}
    \caption{Ablation studies of camera refinement. 
        }
    \label{fig:abl_cam_refine}
    \end{minipage}
\end{figure*}